\documentclass[letterpaper,
english,
reprint,
aps,
prl,
footinbib,
longbibliography,
amssymb,
floatfix]
{revtex4-2}
\usepackage{textcomp}
\usepackage{amsmath}
\usepackage{graphicx}
\usepackage{siunitx}
\usepackage{hyperref}

\usepackage[dvipsnames,svgnames,x11names]{xcolor}
\usepackage[markup=underlined]{changes}
\usepackage{xr}

\definechangesauthor[color=BrickRed]{MV}

\usepackage{color}
\makeatletter

\usepackage{babel}

\makeatother

\begin{document}

\title{Tracing Dirac points of topological surface states by ferromagnetic resonance}

\author{Laura Pietanesi$^{1,a}$, Magdalena Marganska$^2$, Thomas Mayer$^2$, Michael Barth$^2$, Lin Chen$^1$, Ji Zou$^3$, Adrian Weindl$^2$, Alexander Liebig$^{2}$, Rebeca D\'iaz-Pardo$^{1}$, Dhavala Suri$^{1,5}$, Florian Schmid$^{2}$, Franz J. Gießibl$^{2}$, Klaus Richter$^{2}$, Yaroslav Tserkovnayk$^{3}$, Matthias Kronseder$^{2,a}$, Christian H. Back$^{1,4,b}\footnote{christian.back@tum.de}$}

\affiliation{$^1$ School of Natural Sciences, Department of Physics, Technical University of Munich, 85748 Garching, Germany \linebreak
$^2$ Institute for Experimental and Applied Physics, University of Regensburg, 93040 Regensburg, Germany \linebreak
$^3$ Department of Physics and Astronomy, University of California, Los Angeles, California 90095, USA \linebreak
$^4$ Center for Quantum Engineering (ZQE), Technical University Munich, 85748 Garching, Germany \linebreak
$^5$ Center for Nanoscience and Engineering, Indian Institute of Science, Bengaluru 560 012, India \linebreak
$^a$ These authors have contributed equally \linebreak
$^b$ christian.back@tum.de}

\vspace{2cm}
\begin{abstract}
Ferromagnetic resonance is used to reveal features of the buried electronic band structure at interfaces between ferromagnetic metals and topological insulators. By monitoring the evolution of magnetic damping, the application of this method to a hybrid structure consisting of a ferromagnetic layer and a 3D topological insulator reveals a clear fingerprint of the Dirac point and exhibits additional features of the interfacial band structure not otherwise observable. The underlying spin-pumping mechanism is discussed in the framework of dissipation of angular momentum by topological surface states (TSSs). Tuning of the Fermi level within the TSS was verified both by varying the stoichiometry of the topological insulator layer and by electrostatic backgating and the damping values obtained in both cases show a remarkable agreement.
The high energy resolution of this method additionally allows us to resolve the energetic shift of the local Dirac points generated by local variations of the electrostatic potential. Calculations based on the chiral tunneling process naturally occurring in TSS agree well with the experimental results. 
\end{abstract}

\maketitle


\section*{Introduction}
Three dimensional topological insulators (TIs) possess an insulating bulk and metallic topological surface states (TSS) that exhibit spin-momentum locking \cite{RevModPhys.82.3045,moore2010birth}: the carrier spins are orthogonal to their momentum. Around this feature revolves the concept that TSSs could provide a very efficient way to convert spin currents into charge currents and vice versa \cite{zhang2016conversion,Shiomi2014,Deorani2014}. However, since 3D TIs contain heavy elements, the bulk states also exhibit strong spin-orbit coupling and can therefore contribute to spin-charge interconversion (SCI) phenomena through the conventional spin Hall effect \cite{sinova2015spin, mellnik2014spin, yang2016switching, zhu2021highly}.
Angle resolved photoemission electron spectroscopy, ideally with spin analysis, as well as scanning tunneling spectroscopy (STS) are the most popular techniques for the investigation of the electronic band structure with high energy resolution, however, they cannot be applied to study SCI or the buried interface band structure. Methods to study SCI typically require ferromagnet (FM)/TI hybrid structures where the magnetization of the FM can be manipulated via spin-orbit torques generated by the TI or by observing the loss of angular momentum in ferromagnetic resonance (FMR) type experiments.

To realize high-quality hybrid structures with clean interfaces and easily tunable band structure, the method of choice is molecular beam epitaxy (MBE), enabling TI/FM growth without disrupting the vacuum. While MBE growth offers the advantage of well-prepared interfaces, it is associated with significant problems arising from inadvertent doping by crystal defects leading to bulk conducting samples. Therefore, disentangling the contribution of TSSs and bulk conducting states involved in SCI remains an important task. This problem has previously been addressed using concentration doping and the electric field effect to study SCI as well as the spin Seebeck effect \cite{mellnik2014spin,kondou2016fermi,Wang2019, jiang2016enhanced,zhu2021highly}, however, the conversion mechanisms, arising from the bulk or TSS, remain unsettled.
In this work, we combine MBE growth of hybrid TI/FM films with high quality interfaces on substrates suitable for electric backgating with FMR-based spin-pumping experiments which we use as a tool to probe the effective interface band structure of the TI in contact to a FM layer. 
We also propose a theoretical model to describe spin-pumping from FM primarily into TSSs. Electrostatic potential fluctuations arising from crystal defects lead to an imperfect Dirac point alignment throughout the sample surface and give rise to additional features in the spectrum which can be accounted for within our theoretical framework.

\begin{figure*}[t!]
\includegraphics[width = 0.7\textwidth]{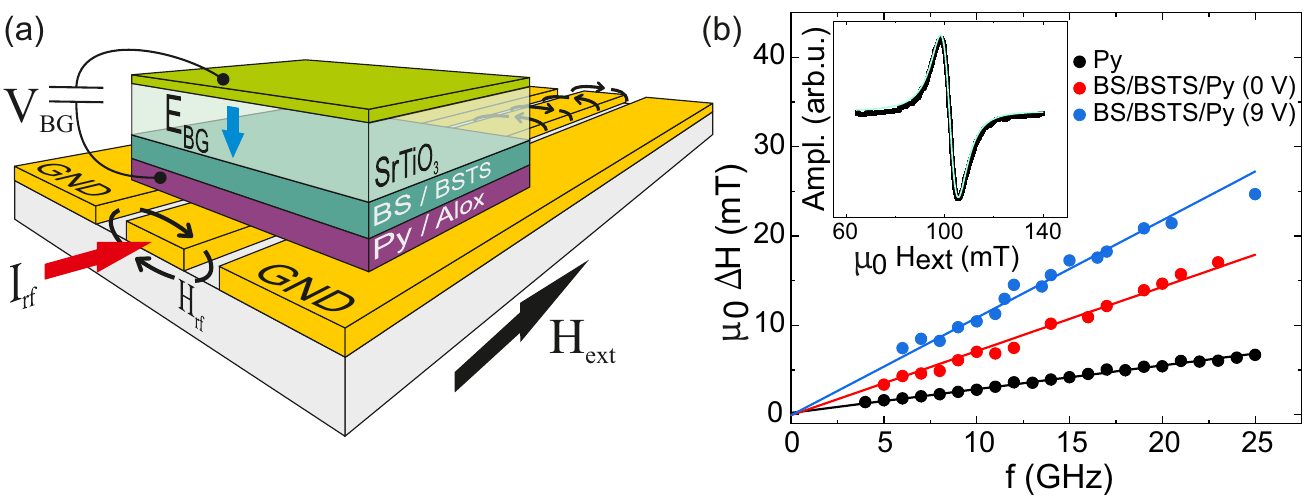}
\caption{(a) Schematic of the spin-pumping FMR setup comprising a coplanar-waveguide (CPW) to externally drive a ferromagnetic system while an external magnetic field $\mathrm{H_{ext}}$ is applied to the system. The temperature of the CPW and sample, which lies face down on the CPW, can be varied between 8-300K. Application of an electric field between the backside and the Py layer in the sample stack SrTiO$_{3}$(111)/(1)BS/(10)BSTS/Py/AlO$_{x}$ is used to gate the system leading to a change of the position of the Fermi level of BSTS. (b) Linewidth dependence on frequency for Py (black curve) at 9~K and (1)BS/(10)BSTS/Py with $x=0.67$ at 10 K and 0 V (red curve) as well as 9 V backgate (blue curve) voltages. The inset shows a typical FMR absorption curve and the corresponding fit (derivative of a Lorentzian line). }\label{fig:Fig1}
\end{figure*}

\section*{Experimental}
The investigated TI heterostructures consist of a single quintuple layer (QL) Bi$_{2}$Se$_{3}$ (BS) and 10 QL (Bi$_{1-x}$Sb$_{x}$)$_{2}$(Te$_{1-y}$Se$_{y}$)$_{3}$ (BSTS) bilayers grown by MBE on SrTiO$_{3}$(111) (STO) substrates. With the n-p heterostructure approach established in \cite{mayer2021transport,Eschbach2015}, we are able to strongly suppress the unwanted carrier contribution due to bulk doping and to determine the position of the Fermi level in the upper TSS ~\cite{danilov2021superlinear}.
The single, intrinsically n-type BS QL is used as an epitaxial and electrostatic seed layer significantly improving the crystal quality \cite{mayer2021transport}. BSTS can be stoichiometrically tuned to p-type: the n-p heterostructure inherently leads to a band bending along the growth direction that allows the Fermi level to be placed well inside the band gap at the top surface \cite{mayer2021transport,danilov2021superlinear}. For details concerning the electrical transport characterization of the complete sample series we refer to section I of the Supplemental Material \cite{SM}. From transport characterization experiments we conclude that in the (1 QL)Bi$_{2}$Se$_{3}$/ (10 QL)(Bi$_{1-x}$Sb$_{x}$)$_{2}$(Te$_{1-y}$Se$_{y}$)$_{3}$ series, samples with $y$ fixed to $\approx$ 0.9 show maximized bulk band gap \cite{Zhi2011,Arakane2012,Xia2013,mayer2021transport}, and $x$ $\approx$ 0.7 places the Fermi energy in the vicinity of the Dirac point at the top surface, as will be explained further below. 
Since spin-pumping is most sensitive to the interface with the ferromagnet, these compositions are ideal for the ferromagnetic resonance/spin-pumping experiments described in the following.

In order to enable FMR measurements, 10 nm thick permalloy (Ni$_{80}$Fe$_{20}$, Py) layers were grown on top of the BSTS layers in a separate MBE chamber connected via an ultra high vacuum tunnel. 7 nm thick AlO$_{x}$ capping layers were grown on top of all samples (with or without the Py layer) to protect the structures from oxidation. Full film samples were placed on top of a coplanar waveguide, as shown in Fig.~\ref{fig:Fig1}(a), and FMR measurements were performed in a temperature range from room temperature to 8 K at fixed frequencies sweeping the external magnetic field. Resonance spectra, as shown in the inset of Fig.~\ref{fig:Fig1}(b) were recorded for frequencies between 5 - 25 GHz, and fitted using the derivative of a Lorentzian profile to extract the resonance field and linewidth. From the frequency dependence of the linewidth, the Gilbert damping parameter $\alpha$, which is a measure for the loss of angular momentum, can be extracted as the slope of a linear fit \cite{Celinski1991,CELINSKI19976}.

\section*{Results}
In TI/FM bilayer structures, spins are pumped from FM into the TI when the magnetization is driven by microwave magnetic fields. Thus, by measuring $\alpha$ and subtracting the intrinsic damping of FM, it is possible to determine the TI's efficiency for absorbing angular momentum near the Fermi energy. The precessing magnetization of the FM layer thereby acts as source of angular momentum and simultaneously as a detector for the loss of angular momentum. By variation of the Sb-content in the BSTS layer or by variation of the backgate (BG) voltage, characteristics of the local band structure and density of states (DOS) in the TI layer can be reconstructed as will be discussed further below.

Figure~\ref{fig:Fig1}(b) shows linewidth vs. frequency obtained from FMR measurements for a Py sample (black curve) and a BS/BSTS/Py sample with $x=0.67$ at 0 V (red curve) and 9 V (blue curve) backgate voltages performed at 10 K. From the linear dependence of the linewidth on the frequency, see Eq. (S1) in the Supplemental Material, we get the following values for $\alpha$: 0.006(9) for pure Py and 0.022(3) for BS/BSTS/Py at 0~V. 
$\Delta \alpha (T) = \alpha_{\mathrm{TI/FM}}(T) - \alpha_{\mathrm{FM}}$(T), indicates the TI's efficiency for absorbing angular momentum as a function of temperature; note that the temperature dependence of $\alpha$ for a pure Py sample grown on STO is almost negligible, see Supplemental Material Fig.~S1(c).

\begin{figure*}[t!]
\centering
\includegraphics[width =0.7\textwidth] {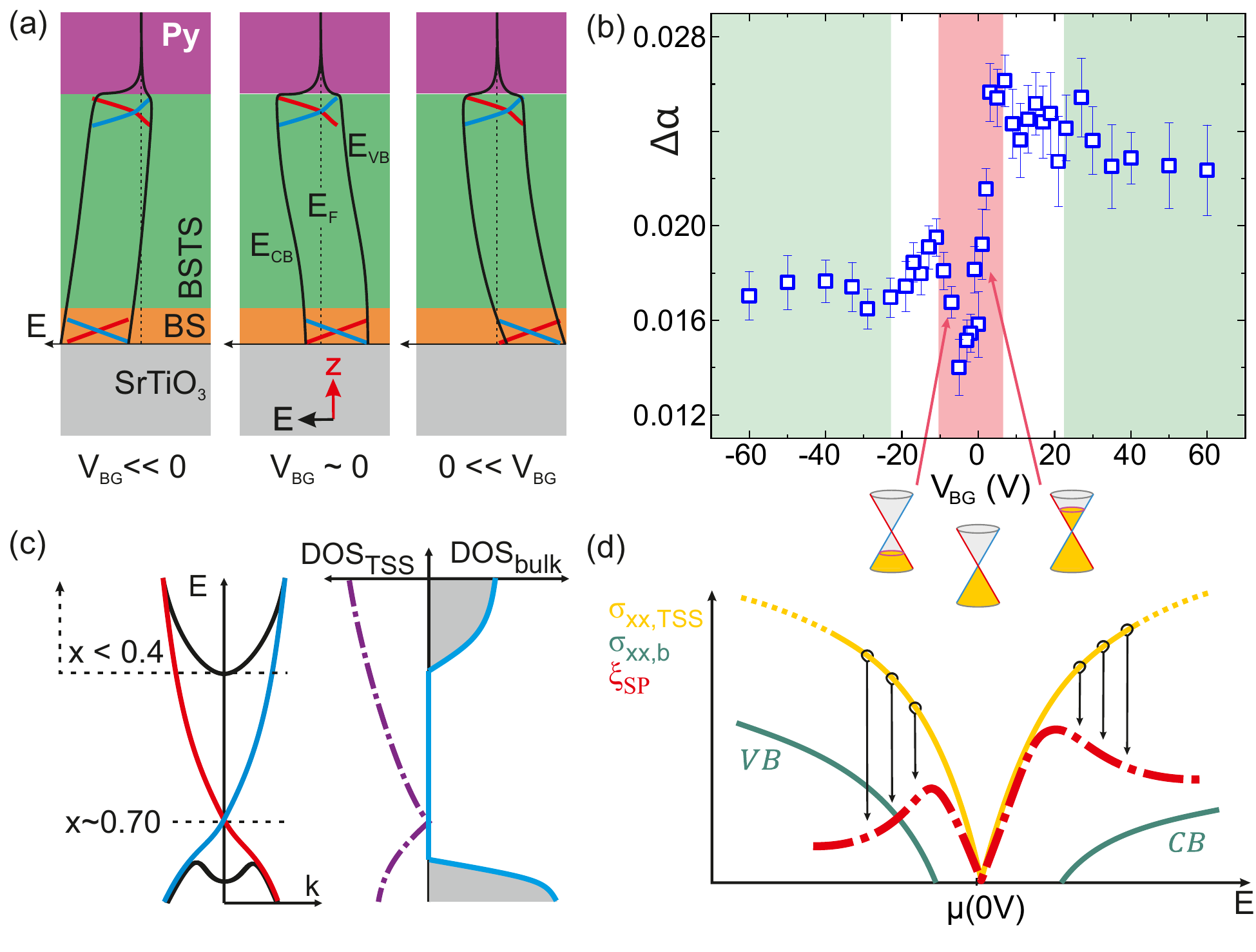}
\caption{\label{fig:Fig2} (a) Schematics of the band diagrams along the growth direction of the heterostructure of BS/BSTS with a Py ferromagnet on top grown on the substrate SrTiO$_3$. At $V_{\mathrm{BG}}\sim0$ the intentional internal band bending sets the Fermi energy in the top-TSS close to the Dirac point. Application of BG voltages distorts the conduction and valence bands changing also the energetic alignment of the top-Dirac cone. (b) Gilbert damping versus applied backgate voltage for a sample with (1)BS/(10)BSTS/Py with $x=0.67$ measured at 10 K. The damping for STO/Py is subtracted: $\Delta \alpha = \alpha_{\mathrm{TI/FM}} - \alpha_{\mathrm{FM}}$. (c) Simplified sketch of the band structure considering bulk states and the Dirac cone at the top surface; the Fermi level position for different Sb concentrations is shown on the left and a sketch of the density of states for TSS and bulk at T=0 is shown on the right. (d) Schematic representation of different contributions of the bulk, $\sigma_{xx,\mathrm{b}}$ (green line), and of the TSSs, $\sigma_{xx,\mathrm{TSS}}$ (yellow line), to the longitudinal conductivity $\sigma_{xx}$ resulting in a spin-pumping efficiency $\xi_{\mathrm{SP}}$ (red dashed-dotted line).  }
\end{figure*}

To investigate $\Delta \alpha$ as a function of the position of the Fermi level, backgated FMR/spin-pumping experiments were performed. 
Efficient backgating is enabled by building a capacitor with a conducting layer at the back side of the STO substrate and using Py as a front electrode. The choice of STO as substrate allows efficient gating, thanks to its structural phase transition at 105 K \cite{WEAVER1959274}: below this temperature the substrate's dielectric constant increases dramatically to around 7000 at $\sim$10K (see Supplemental Material Fig.~S4).  A schematic band diagram as shown in Fig.~\ref{fig:Fig2}(a) middle panel can be drawn for the n-p heterostructure: when the BSTS is p-type and thick enough, the band bending generated by the n-p junction enables a precise control of the Fermi level position in the bulk band gap at the top surface and drastically reduces the metallic contribution of the 1QL BS \cite{mayer2021transport,danilov2021superlinear}. Applying an electric field via backgating allows shifting the Fermi level primarily at the interface to the STO substrate. Since the band diagram is anchored at the metallic Py interface, the band diagram in growth direction can be viewed as a lever arm when applying electric fields. Consequently, a small non-zero shift of the band diagram in the top layers will result when backgating the heterostructure, see Fig.~\ref{fig:Fig2}(a). This enables extremely precise control of the position of the Dirac point of the top-TSS with respect to the Fermi level with a shift of $\frac{dE_{\mathrm{DP}}}{dV_{\mathrm{BG}}}= -3 \frac{meV}{V}$ as verified in STM/STS experiments (see Supplemental Material Fig.~S5). Note that since the energy of the precessing magnetization is in the range of a few tens of $\mu$eV for the frequency range used in our experiments, FMR/spin-pumping has the necessary high energy resolution to resolve features within the band gap.
To determine the optimum stoichiometry for the spin-pumping investigations, the temperature dependence of $\Delta \alpha$ has been measured for the sample series with $y$ fixed to $\approx$ 0.90 and variable Sb concentration $x$, see Supplemental Material Fig.~S1. 
We would like to emphasize here that spin-pumping is mostly sensitive to the FM/TI interface. In contrast, transport experiments reflect the whole conducting layer consisting of bottom-, top-TSS and bulk channels \cite{Just2020,Backes2017}. Thus, for choosing the most suitable FM/TI-hybrid either temperature dependent spin-pumping experiments or other methods like the photogalvanic effect \cite{danilov2021superlinear} can be used. For the former approach, $\alpha(T)$ reflects the temperature dependence of the top-TSS and parts of the bulk systems, see Fig.~S2, and can be summarized as follows: when the Fermi energy in the topmost TI-layers is within the bulk band gap, see Fig.~\ref{fig:Fig2}(a) middle panel and Fig.~\ref{fig:Fig2}(c), the TSS governs the efficiency of spin-pumping at low temperatures and consequently $\Delta \alpha$ starts to increase below $\approx 70\,$K. At elevated temperatures, thermally populated bulk states counteract this effect \cite{Skinner2012,Skinner2013}, leading to reduced dissipation of angular momentum into the TSS. For the latter method, it was shown in \cite{danilov2021superlinear} for the same TI-structure that the Fermi energy is indeed close to the Dirac point in the top-TSS for $x\approx 0.7$. Hence, both studies confirmed the choice of $0.67 < x < 0.73$ as the ideal concentration range even when replacing the AlO$_x$ capping layer with Py. Please note, the Fermi level/Dirac point positions determined by ARPES \cite{mayer2021transport} and STM (see section IV of the Supplemental Material) on uncapped films most likely differ from the positions in the Py/AlO$_x$ capped samples measured in the FMR experiments. Removal of the Se capping layers leaves sample surfaces that are likely to be n-type due to residues and Se vacancies, as they are generally n-type dopants for Bi-based TIs.

Subsequently, in a first overview experiment, $\Delta \alpha$ was measured as a function of backgate voltage in the range of $\pm$60 V, see Fig. \ref{fig:Fig2}(b), for a sample with $x=0.67$ at 10~K. We observe that $\alpha$ can be significantly manipulated by backgating; as shown exemplarily in Fig.~\ref{fig:Fig1}(b), the value of $\Delta \alpha(V) $ increases from 0.015(3) at 0~V to 0.024(3) for a backgate voltage of +9~V.  The results summarized in Fig.~\ref{fig:Fig2}(b) can be divided into three regions: two saturation regimes for large backgate voltages (shaded in green) and a region around zero backgate voltage (shaded in red). In the red shaded region a minimum of $\Delta\alpha$ is observed at -5 V. Away from the Dirac point, but still in the red region, $\Delta\alpha$ increases, followed by a decrease until it reaches saturation in the green shaded region. At large positive or negative gate voltages, the Fermi energy in the layers close to the STO substrate touches the bulk conduction or valence bands, respectively, see Fig. \ref{fig:Fig2}(a). This means that the bottom layers become highly conductive and screen any further changes to the electric backgate; consequently saturation of $\Delta\alpha$ is reached. It is important to note that due to this effect and the uncertainty in the electrostatic lever arm from bottom to top layer in the np-heterostructure, the linear gate voltage scale does not translate into a linear energy scale for the position of the Fermi level in the top-TSS.

\begin{figure}[t!]
	\centering
	\includegraphics[width =0.9\columnwidth] {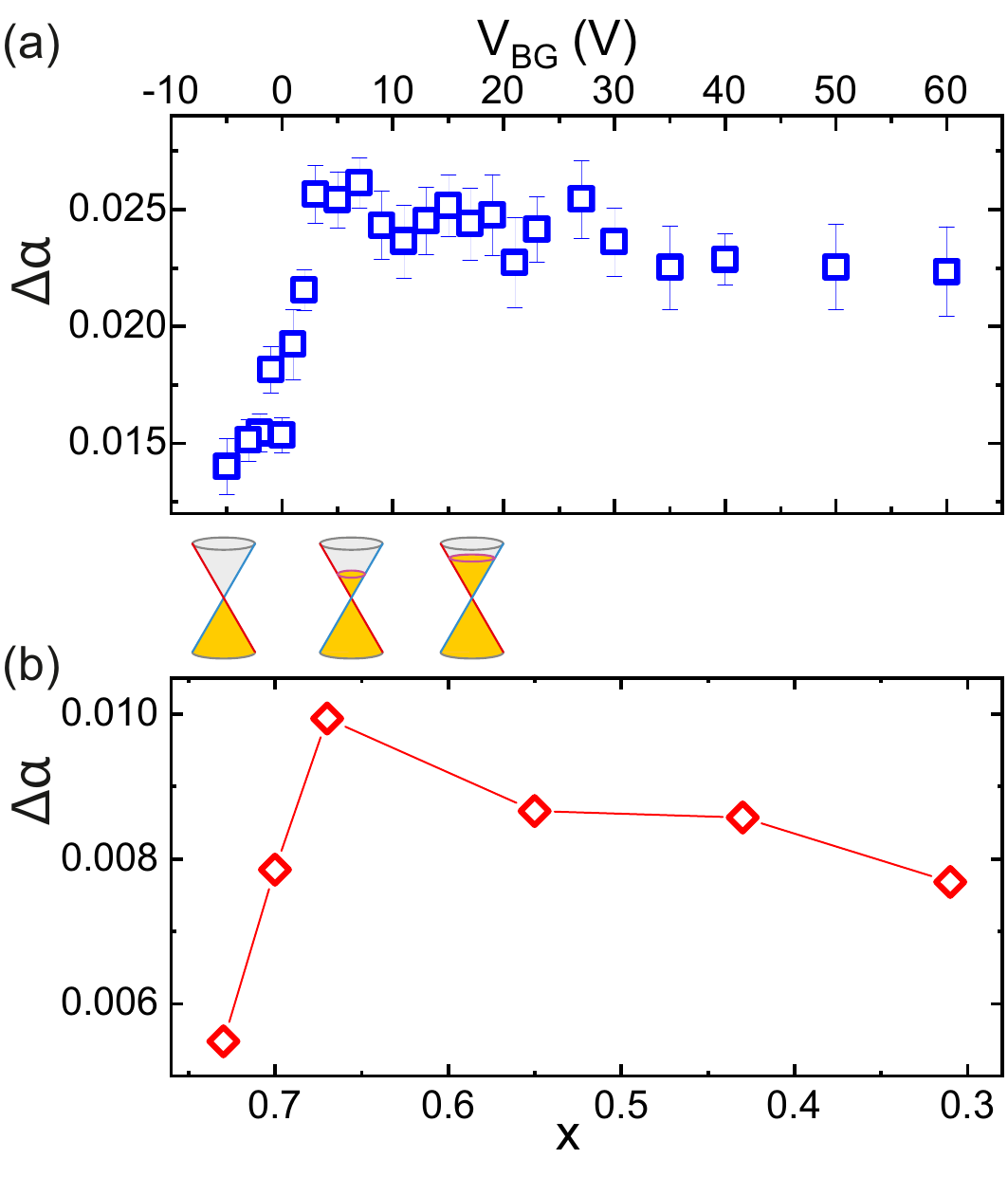}
	\caption{\label{fig:Fig3} To enable a direct comparison, a part of the data presented already in Fig.~\ref{fig:Fig2}(b) is replotted in (a) for the sample with $x\sim$0.67 measured at 10 K. At $V_\mathrm{BG}\sim$-5V the Fermi energy is at the Dirac point. (b) $\Delta\alpha$ dependence on Sb concentration measured at 10 K at $V_{BG} = 0$. The Fermi energy shifts towards the conduction band with increasing $V_\mathrm{BG}$ or decreasing $x$ as shown in the sketch of the Dirac cone between the panels. The measurements shown in (a) have been recorded on the same sample as shown in (b) at x=0.67 with 12 months separation.}
\end{figure}

We were able to show a reciprocity between the two different experiments performed: varying the Sb concentration and measuring $\Delta \alpha$ on different samples at low T agrees well with the backgated measurements of $\Delta \alpha$ on single samples, as shown in Fig.~\ref{fig:Fig3}. Note that in these experiments the relative position of the Fermi energy in the top-TSS was varied in two ways, by changing the stoichiometry via Sb-content $x$ and by application of a backgate voltage $V_{\mathrm{BG}}$. Both dependencies, $\alpha(E_F(x))$ and $\alpha(E_F(V_{\mathrm{BG}}))$, show very similar behavior and are juxtaposed in Fig.~\ref{fig:Fig3}. This is a first hint that the DOS of the TSS governs spin-charge conversion and therefore the loss of angular momentum measured with $\Delta \alpha$. The agreement between the two measurements is a further confirmation that the electrostatic backgate is an efficient way of varying the Fermi energy. Note that while a variation of SCI within the band gap disagrees with the prediction of being approximately constant as suggested in \cite{Wang2019}, it is fully in line with the results shown in \cite{kondou2016fermi} and the following theoretical considerations.

\section*{Theoretical Background}
\begin{figure*}[t!]
\centering
\includegraphics[width =1\textwidth]{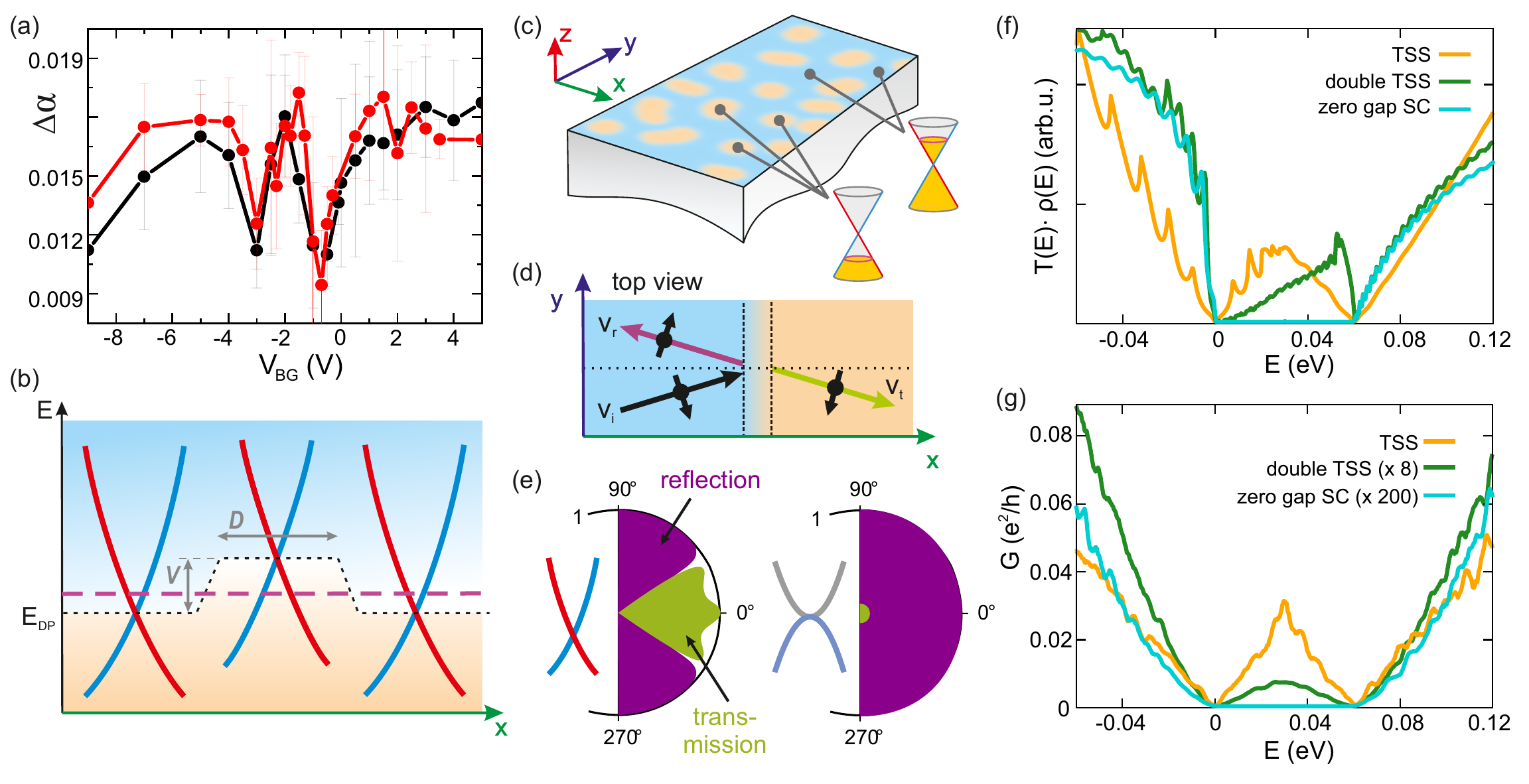}
\caption{\label{fig:Fig4} (a) High-resolution scan of $\Delta\alpha$ around $V_{\mathrm{BG}}\sim 0$. Both measurements (red and black dots) have been performed on the same sample with (1)BS/(10)BSTS/Py and $x=0.7$, measured at T=10K. (b) Band alignment for a spatially extended potential barrier in a single TSS. (c) Corrugated potential energy landscape with local n-p junctions for $E_\mathrm{F}$ between the Dirac points (purple dashed line in (b)). A right moving electron with spin up which crosses the n-p boundary has to become a right moving hole with spin up ((b)), undergoing the chiral tunneling and reflection process as shown in (d). In chiral systems some incidence angles lead to enhanced transmission ((e) left), for a non-chiral zero-gap semiconducting system transmission is exponentially suppressed (right). (f) Analytically calculated quantum mechanical transmission for three systems (single TSS, double TSS and zero gap semiconductor) with a potential barrier (as in (b)) of height $V=60$~meV and width $D=60$~nm, weighted by the DOS. The chiral systems, single and double TSS, show two dips at the Dirac point energies with a pronounced maximum between them. Removing the chiral nature of the system results in zero transmission for energies at which n-p or p-n junctions occur as in this case the transmission is mediated by evanescent waves. (g) Numerical transport calculations for a single potential step in the same systems as in (f) at $T = 10$~K reveal the same qualitative features as the quantum mechanical approach in (f). The results in (f) and (g) are rescaled for visibility (rescaling factor in brackets in the legend, see also the Supplemental Material). }
\end{figure*}

In order to understand how the spin-pumping signal is related to the band structure and density of states of the TI, the theory of spin-pumping from FM into a TI has to be examined. Let us assume the injection layer Py to be magnetized along the $y$-direction, with the Cartesian directions schematically indicated in Fig.~\ref{fig:Fig4}(c). Then, angular momentum transfer is related to the transformation of a spin current directed along the $z$-direction into an in-plane charge current in the TI. The TI/FM heterostructure can be described by a minimal Hamiltonian \cite{Tserkovnyak2012} $H = v_\mathrm{F} (\mathbf{p}-e\mathbf{A})\cdot \mathbf{z}\times\hat{\mathbf{\sigma}} + J \,m_z \hat{\sigma}_z$ including the kinetic energy of Dirac electrons and a term describing the coupling $J$ of the Dirac electron spin $\hat{\mathbf{\sigma}}$ to the spin dynamics $\mathbf{m}(t)$ of the ferromagnet. The in-plane components of the coupling between the FM and the TSS are absorbed into the vector potential $\mathbf{A}$: $\mathbf{A} = \frac{J}{ev_\mathrm{F}} \mathbf{m}\times\mathbf{z} = \frac{J}{ev_\mathrm{F}}(m_y, -m_x,0)$. 
If the static in-plane magnetization points along the $y$-direction, the external drive with frequency $\omega$ will lead to an elliptical precessional motion with amplitude $\delta m$ in the $x$-direction and a $z$-component of the precession which is much reduced due to the demagnetizing field. At ferromagnetic resonance the vector potential gives rise to an in-plane electric field $\mathbf{E}(t) = -\partial_t \mathbf{A} = -\frac{J \delta m \: \omega}{e v_\mathrm{F}} \sin \omega t \: \mathbf{y}$.
This in-plane electric field along the $y$-direction eventually drives an electric current and dissipates power according to 
\begin{equation}\label{powerjoule}
    P\equiv  \mathbf{j}\cdot\mathbf{E} = \sigma  E^2_y  = \sigma  \left(\frac{J \delta m \: \omega}{e v_\mathrm{F}}\right)^2\, ,
\end{equation} 
where $P$ is the power density and 
$\mathbf{j}$ the current density. If we assume an isotropic in-plane longitudinal conductivity $\sigma \equiv \sigma_{xx} = \sigma_{yy}$, this equation is valid for different orientations of the FM. On the other hand, the rate of energy dissipation per unit volume, 
associated with the predominantly in-plane spin dynamics, can be evaluated to 
\begin{equation}\label{poweralpha}
    P = \frac{\alpha s}{2} \dot m_x^2 = \alpha \, \frac{ s \, \delta m^2 \, \omega^2}{2}  \, ,
\end{equation}
where $\alpha$ is the Gilbert damping parameter and $s$ is the local spin density. Combining (\ref{powerjoule}) and (\ref{poweralpha}) leads to 
\begin{equation}\label{alphasigma}
    \alpha = \frac{2 J^2}{s e^2 v_\mathrm{F}^2} \sigma \, .
\end{equation}
Hence, for magnetic coupling to 2D-TSS, damping is proportional to the longitudinal conductivity of the Dirac electrons: $\alpha \propto \sigma_{xx, \mathrm{TSS}}$ \cite{PhysRevB.89.165307,PhysRevB.95.094428,Tserkovnyak2012}, assuming predominantly in-plane spin fluctuations. Note that this dependence is rooted in the form of the kinetic term in the Hamiltonian leading to the electron velocity operator $\hat{\mathbf{v}} = v_\mathrm{F} \, \mathbf{z}\times \hat{\mathbf{\sigma}}$, which is essentially given by the spin.

Although 180$^\circ$-backscattering is strictly forbidden for TSS, all other angles are generally allowed for momentum scattering processes. In real systems any source of disorder leads to scattering and the type of disorder sets, in general, a finite scale for the allowed momentum transfer ($|\Delta \mathbf{k}| < k_0$, with $k_0$ being a cutoff for the maximum momentum transfer). Away from the Dirac point ($k_\mathrm{F} > k_0$), the allowed scattering angles are limited by this cutoff which increases the conductivity.  Close to the Dirac point, scattering covers the whole Fermi surface ($k_\mathrm{F} < k_0$), albeit anisotropically, due to the Dirac suppression of the backscattering. This leads to a reduced conductivity at low doping. In sec. VI of the Supplemental Material, the influence of non-isotropic scattering on the conductivity is further investigated.

Translating the longitudinal conductivity to magnetic damping leads to the following scenario: within the band gap, $\alpha$ is fully governed by the TSS, see Supplemental Material Fig.~S2. It is thus necessarily low at the Dirac point and increases away from the Dirac point as shown in Fig.~\ref{fig:Fig2}(b). When the band structure is richer, i.e. when additionally considering bulk states outside the band gap (with conductivity $\sigma_{xx,\mathrm{b}}$), or trivial in-gap states (e.g. impurity levels or interface states with conductivity $\sigma_{xx,\mathrm{i}}$), $\alpha$ is dictated by the interplay of different conductivities. Both, $\sigma_{xx,\mathrm{b}}$ and $\sigma_{xx,\mathrm{i}}$ have the effect of lowering the conductivity $\sigma_{xx,\mathrm{TSS}}$ due to hybridization and consequently enhanced scattering probabilities. This translates into a simple recipe for the spin-pumping efficiency: the larger the conductivities of trivial states, the lower the spin-pumping efficiency. A simple version of these arguments considering only TSS and bulk states is sketched in Fig.~\ref{fig:Fig2}(d), with $\sigma_{xx,\mathrm{TSS}}$ governing the overall spin-pumping efficiency $\xi_{\mathrm{SP}}$ and $\sigma_{xx,\mathrm{b}}$ lowering $\xi_{\mathrm{SP}}$. As the valence band edge is closer and the valence band conductivity is larger than the conduction band edge and conductivity \cite{Wang2019,mellnik2014spin} the reduction of $\alpha$ sets in earlier and is stronger towards the valence band ($E<0$) than in the case of the conduction band ($0<E$), meaning that the total damping remains at a larger value when the Fermi energy moves towards the conduction band compared to the valence band direction. The red curve in Fig.~\ref{fig:Fig2}(d), representing the expected spin-pumping efficiency, agrees with the measured data shown in Fig.~\ref{fig:Fig2}(b) for $x=0.67$.

\section*{High-resolution scans}
So far, in the theoretical description of the spin-pumping mechanism, we assumed a homogeneous system, i.e. a system where the band structure is independent of the lateral position. However, in real extended systems the energetic position of the Dirac point with respect to $E_\mathrm{F}$ varies as a function of lateral position \cite{Knispel2017,Bomerich2017,mann2013mapping} resulting in p- and n-type regions when $E_\mathrm{F}$ is tuned to the vicinity of the Dirac points, see Fig.~\ref{fig:Fig4}(b),(c). As a consequence, the spin-pumping efficiency must also vary locally, i.e. $\xi=\xi(\mathbf{r})$, with $\mathbf{r}$ being a lateral position in the 2D-TSS. To detect these variations experimentally, we performed high-resolution backgate voltage scans near the energetic position of the Dirac point. This resulted in the appearance of two distinct minima in the dependence of magnetic damping on backgate voltage, which will be explained below. Indeed, two distinct minima could be reproducibly measured as shown in Fig.~\ref{fig:Fig4}(a).

Spin-pumping integrates over the potential landscape of the entire sample near the interface between FM and the TI, and thus two scenarios can be drawn if local variations are taken into account. In the first scenario, which is in the limit of a TSS with low overall conductivity, the spin-pumping efficiency is fully determined by the instant process of spin-charge interconversion. In this limit the actual charge transport within the TSS plays a minor role. 
As an example, if we assume a corrugated energy landscape with two distinct potential levels randomly distributed across the surface, the integrated spin-pumping efficiency will have two minima at the two Dirac point energies. The relative depth of the two minima reflects the ratio of the areas occupied by the two different potential levels. Away from the Dirac points the potential variations become negligible. In \cite{mann2013mapping} $\mathrm{Bi_{Se}}$-antisite defects lead to the occurrence of two energetically distinct defect states depending on the position of the defect in the lattice. As a result, potential fluctuations in the Dirac point energy centered around two main energies were observed in \cite{mann2013mapping} and are in line with our observation of two main dips in damping.

In the second scenario the overall TSS conductivity is assumed to be large and charge transport within the TSS after the SCI process does play an important role. Considering again a corrugated energy landscape with a global Fermi energy in the vicinity of the Dirac points, as in Fig.~\ref{fig:Fig4}(b) with n- and p-type regions, an additional transport mechanism occurs, that of chiral tunneling.

The chiral nature of massless Dirac fermion systems, like graphene \cite{Young2009,Stander2009,Martin2008} and TIs, is the basis for this positive contribution to the conductivity between the two Dirac point levels and is closely related to Klein tunneling. The original Klein tunneling describes the angle-dependent tunneling process of a Dirac fermion propagating across a very high electrostatic barrier which leads in non-relativistic systems to an exponential decay of the state, but in relativistic cases the transmission probability reaches unity for normal incidence, independent of the barrier height and width \cite{Klein1929,NgocHan2016,Habib2015,Xie2017,Richter2012,Katsnelson2006}. 
In chiral systems, such as the spin-momentum locked TSS, charge carriers propagate through the barriers not via evanescent waves as usual, but via travelling waves belonging to the other branch of states. In the n-p-n tunneling depicted in Fig.~\ref{fig:Fig4}(b), the incoming electrons from the positive Dirac cone (n) are transmitted across the central barrier through the extended eigenstates from the negative Dirac cone (p). The overall transport is determined by the interplay between the momentum selection rules, spinorial compatibility between incoming and transmitted states, and the density of states of the incoming carriers. 
The Dirac cone of the TSS has two special features -- it is chiral and its density of states (DOS) vanishes at the Dirac point. In order to gauge the importance of each of these features, we compare the TSS to two other systems. One of them consists of two coupled TSSs, analogous to bilayer graphene, and is chiral but always has finite DOS. The second is a zero-gap semiconductor, which is not chiral, but has constant DOS. Both these artificial systems are discussed purely for the sake of illustration. Note that in the following description and calculation bulk contributions are neglected. Now let us consider the transmission through a 1D barrier extending in the $y$ direction, such as shown in Fig.~\ref{fig:Fig4}(b). For a general angle, the electrons are partly reflected and partly transmitted (see Fig.~\ref{fig:Fig4}(e)). At normal incidence, transmission reaches unity for the single TSS, while for the double TSS it is strictly zero and the charge is transmitted only for oblique incidence angles. For the zero-gap semiconductor transmission is exponentially suppressed, as shown schematically in Fig.~\ref{fig:Fig4}(e). When we integrate the quantum mechanical transmission over the incidence angles, and multiply it by the density of states of the incoming electrons, for both chiral systems we find an enhanced transmission in the energy range between two minima corresponding to the top and bottom of the barrier. In the same energy range, transmission for the non-chiral zero-gap semiconductor is fully suppressed as can be seen in Fig.~\ref{fig:Fig4}(f). Similar conclusions can be drawn from a numerical calculation of transport in the three systems with the same potential landscape (cf. Fig.~\ref{fig:Fig4}(g)), where for the single and double TSS we see again a conductance peak within the barrier, flanked by two dips, while for a zero gap semiconductor the conductance in this energy range is zero. Qualitatively, the same results are obtained numerically for a potential landscape with 2D potential islands, shown in Sec. VIII of the Supplemental Material. The central conductance peak ﬂanked by two dips persists also for 2D potential disorder. The source of the discrepancy between the shapes of the calculated and measured curves is the highly nonlinear coupling between the gate voltage and the position of $E_\mathrm{F}$ of the top-TSS in our devices, hence the true shape of the conductance, $G(E_\mathrm{F})$, inferred from spin-pumping, may be different. For a more detailed discussion of chiral tunneling, modelling of the three systems and transport calculations we refer the reader to Sec. VII and VIII of the Supplemental Material.

Summarizing the second scenario in which after the occurrence of SCI further charge transport through a corrugated energy landscape plays a role: Analytical and numerical calculations confirm the suppression of the conductivity exactly at the Dirac points, leading to a similar result as in the first scenario with the main difference that in this case the levels of the minima are independent of the area ratio as long as the percolation limit is not reached. Furthermore, charge transport requires passing both np- or pn-boundary at least once within the mean free path. All calculations confirm that the formation of a conductivity peak (and hence a peak in the spin pumping efficiency) between two distinct dips is caused by the chiral nature of the system.

Finally, for the sake of completeness, we would like to mention a third possible scenario which, however, we believe is rather unlikely. It is based on the possibility of electron hopping between the FM and the TSS \cite{Arimoto2021}. In \cite{Arimoto2021} it has been shown that due to the hybridization between the electronic bands of the TSS and a mono-crystalline FM, for strong enough coupling and a precise matching of the bands of the FM and the TI avoided crossings can occur in the band structure, opening smaller gaps within the main gap of the TI. This in turn may lead to the occurrence of several minima in the conductivity and thus in spin-pumping efficiency. Our polycrystalline permalloy films contain grains with many lattice orientations, each of them potentially hybridizing with different parts of the TSS Dirac cone. In consequence, we would not expect to observe well-defined dips/gaps, but rather a background uniform in energy.

\section*{Conclusion}
In conclusion, we have shown that spin-pumping in combination with backgating can be used as a high resolution tool to reveal details of the buried energy landscape of topological insulators. The sensitivity of the spin-pumping mechanism to features in the TI band structure was theoretically explained by the dependence of the damping parameter on the longitudinal conductivity. We also measured the dependence of $\Delta\alpha$ on the Fermi energy in two equivalent ways: changing the stoichiometry and applying a backgate. Both measurements are in good agreement and show the exceptional control of the position of the Fermi energy in the top-TSS of the n-p-heterostructure.
Further, application of this method with high resolution shows a rich structure of the spin-pumping efficiency within the bandgap which can be related to the corrugated energy landscape within the TI. Among the three possible explanations for these extra features, the chiral tunneling effect seems to be most likely since it does not rely on low conductivity of the TSS or a specific matching of the lattice and electronic band structure of FM and TI. In addition, only the chiral tunneling effect can lead to the formation of a peak between conduction dips. On the contrary, in non-chiral systems the conductivity would be fully suppressed for energies within the range of potential fluctuations. 

\section*{Acknowledgement}
We acknowledge the financial support of the Deutsche Forschungsgemeinschaft through Project ID 422 314695032-SFB1277 (Subprojects A01, A07, A08 \& B04). The work at UCLA (JZ and YT) was supported by the US Department of Energy, Office of Basic Energy Sciences under Grant No. DE-SC0012190.




%

\end{document}